\documentclass[seceqn]{elsart}

\usepackage{graphicx}
\usepackage{arydshln}

\newcommand{\eq}[1]{(\ref{#1})}

\begin{document}

\begin{frontmatter}

\title{Non-equilibrium thermodynamics and fluctuations}

\author{R. van Zon and E.G.D. Cohen}
\address{The Rockefeller University, 1230 York Avenue, New York 10021, USA}
\date{5 November 2003}

\begin{abstract}
In the last ten years, a number of ``Conventional Fluctuation
Theorems'' have been derived for systems with deterministic or
stochastic dynamics, in a transient or in a non-equilibrium stationary
state. These theorems gave explicit expressions for the ratio of the
probability to find the system with a certain value of entropy (or
heat) production to that of finding the opposite value. A similar
theorem for the fluctuations of the \emph{work} done on a system has
recently been demonstrated experimentally for a simple system in a
transient state, consisting of a Brownian particle in water, confined
by a moving harmonic potential. In this paper we show that because of
the interaction between the stochastic motion of the particle in water
and its deterministic motion in the potential, very different new
\emph{heat} theorems are found than in the conventional case. One of
the consequences of these new heat Fluctuation Theorems is that the
ratio of the probability for the Brownian particle to absorb heat from
rather than supply heat to the water is much larger than in the
Conventional Fluctuation Theorems. This could be of relevance for
micro/nano-technology.
\end{abstract}

\begin{keyword}Fluctuation Theorems, Thermodynamics, Brownian motion.\PACS 05.40.-a \sep 05.70.-a \sep 44.05.+e \sep 02.50.-r\end{keyword}

\end{frontmatter}

\section{Introduction}
\label{intro}

As is well known the First and Second Law of Thermodynamics only
involve averages of the physical quantities of macroscopic systems but
say nothing about their fluctuations. In particular the Second Law for
Irreversible Processes states that the average\footnote{Here, as well
as in the rest of the paper, a bar means an average.} heat
${\overline{Q}}$ internally produced in an irreversible process has to
be positive. In the last ten years a number of fluctuation theorems
have been derived for the fluctuations of thermodynamic properties in
non-equilibrium stationary \cite{Evansetal93,GallavottiCohen95a}, as
well as transient states \cite{Evansetal94,CohenGallavotti99}, which
constitute refinements of the laws of thermodynamics in so far that
they take into account a property of their fluctuations which goes far
beyond the statement that ${\overline{Q}} > 0$. In this respect a
generalization of thermodynamics including fluctuations has been in
progress.

It should be pointed out that these new fluctuation properties are
valid for large fluctuations around a non-equilibrium stationary state
possibly far from equilibrium. As such they differ from fluctuations
dealt with in the context of Irreversible Thermodynamics, such as the
Fluctuation Dissipation Theorem and the Onsager relations (as treated
e.g.  in the classical book by De Groot and Mazur
\cite{DeGrootMazur}), which refer to small fluctuations around
equilibrium.

The Fluctuation Theorems (FT) will be divided into two classes:
Conventional (CFT) and New (NFT) Fluctuation Theorems, where in the
latter case the word ``theorem'' is premature. We will not confine
ourselves here to heat fluctuations alone but also consider work and
energy fluctuations. In the literature until now the overwhelming
number of papers has dealt exclusively with heat in the form of
entropy production as it occurs in the Second Law.

\subsection{Conventional FT}
\label{seccft}

The first discovery of a FT of the kind we will discuss here was
mainly numerical in a computer simulation by Evans, Cohen and Morris
in 1993 \cite{Evansetal93}.  This FT for a non-equilibrium stationary
state was inspired by dynamical systems theory notions
\cite{GallavottiCohen95a}.  A similar FT for a transient state was
formulated in 1994 by Evans and Searles \cite{Evansetal94}. In this
paper we will restrict ourselves to the non-equilibrium Stationary
State Fluctuation Theorem rather than to the Transient Fluctuation
Theorem.

The CFT deals with the fluctuations of work as well as heat in a
finite dynamical system.  The many particle (Hamiltonian) system is
subject to an external force, which does work on the system.  However,
this would heat up the system if it were not for an ingenious internal
thermostat, realized by adding a damping term to the equations of
motion.  The dynamics of this system is purely \emph{deterministic}.
The dissipation taking place in the system is manifested in a
contraction of the accessible phase space of the system which can be
related to a (generalized) physical entropy production in the
system\footnote{The dynamical systems are required to have an
isoenergetic Gaussian thermostat for this to be strictly true,
otherwise correction terms appear, though these might possibly vanish
in the large time or system size limit.}. A FT was derived under a
number of assumptions for such a system for the heat or entropy
production, which will be given below.  Some years later, Kurchan (in
1998) \cite{Kurchan98} and Lebowitz and Spohn (in 1999)
\cite{LebowitzSpohn99} derived a similar FT for a system with purely
\emph{stochastic} dynamics under a number of~assumptions.

In both cases a CFT was found, which can be written in the form:
\begin{equation}
\frac{P(Q_\tau)}{P(-Q_\tau)} \; \; 
\:\raisebox{-5pt}{\overrightarrow{{\tau \rightarrow
\infty}}}\: \; e^{\beta Q_\tau}
\label{1}
\end{equation}
Here $P(Q_\tau)$ is the probability that the fluctuating heat
\emph{produced} in the system during a time $\tau$ has a value
$Q_\tau$.  Therefore, $-Q_\tau$ is a value of the fluctuating heat
\emph{absorbed} by the system of the same magnitude (during an equally
long time).  In taking the limit $\tau\to\infty$ in eq.~\eq{1},
$Q_\tau\sim\tau$ scales as $\tau$.  The $Q_\tau$ (and later $W_\tau$)
in the stationary state can be visualized as fluctuations on segments
of duration $\tau$, obtained by cutting a very long stationary state
trajectory of the system in phase space into segments.

For systems with deterministic dynamics the {\emph{same}} CFT holds
for the work done in a time $\tau$ denoted by $W_\tau$, i.e., $Q_\tau$
in eq.~\eq{1} can be replaced by $W_\tau$.  The reason is that
$Q_\tau$ and $W_\tau$ are represented by the same mathematical
expression here, due to the fact that the thermostat converts all
external work done on the system into internal heat.

There is a connection between this CFT and theorems discussed in the
context of Irreversible Thermodynamics. Gallavotti proved for
deterministic dynamical systems, that near equilibrium, eq.~\eq{1}
leads to the Onsager relations, the Fluctuation Dissipation Theorem
and the Green-Kubo relations for the transport coefficients
\cite{Gallavotti96}.  This seems to imply that the fluctuations
incorporated in eq.~\eq{1} go beyond Irreversible Thermodynamics,
i.e., beyond the linear, near equilibrium regime.  There is no
estimate available for the range of validity of Irreversible
Thermodynamics nor are there any results in this nonlinear regime to
date.

\subsection{The system}
\label{secsystem}

\begin{figure}[t]
\centerline{\includegraphics[height=0.25\textheight]{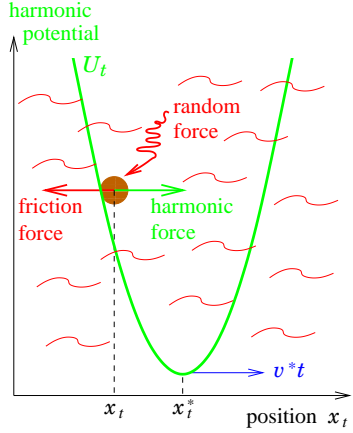}}
\caption{Brownian particle in a moving harmonic
potential (cf. ref.~\cite{Wangetal02}).}
\label{fig1}
\end{figure}

We are interested in this paper in a quasi-many particle system: a
Brownian particle suspended in water and restricted in its motion by a
laser induced harmonic potential, which is pulled through the water
with a constant velocity~$\mathbf v^*$ (cf. fig. \ref{fig1}). This
system was introduced earlier in a somewhat different context by Wang
{\it et al.} in 2002 \cite{Wangetal02}.

Contrary to the systems in section~\ref{seccft}, work and heat are not
identical for this system. The fluctuations of the work done on it as
well as of the heat produced by it in a time $\tau$ were computed
based on an overdamped Langevin equation
\cite{VanZonCohen02b,VanZonCohen03a}. Although this system is in
principle a many particle system, the many degrees of freedom of the
water have been contracted to those of the (Stokes) friction of the
Brownian particle and the strength of its assumed Gaussian white
noise.  In dimensionless units,\footnote{Compared to
ref.~\cite{VanZonCohen02b}, this means $k=\alpha=k_BT=1$.} the
Langevin equation reads
\begin{equation}
0 = - \dot{\mathbf x}_t - (\mathbf x_t - \mathbf x^*_t) + \mathbf\zeta_t .
\label{2}
\end{equation}
Here the first term on the right hand side  represents the
friction of the Brownian particle in the water, the second term represents
the harmonic force due to the harmonic potential
\begin{equation}
U_t(\mathbf x_t) = \frac{1}{2} (\mathbf x_t - \mathbf x^*_t)^2,
\label{3}
\end{equation}
while the third term represents the fluctuations of the Brownian
particle due to the thermal motion of the water molecules. $\mathbf
x_t$ and $\mathbf x^*_t$ are the positions of the Brownian particle
and the minimum of the potential at time $t$, respectively (cf.\
fig.~\ref{fig1}). Essential is that contrary to the pure dynamics of
the two previous systems, satisfying the CFT, here a {\emph{mixed}}
dynamics occurs: {\emph{deterministic}} due to the harmonic force and
{\emph{stochastic}} due to the water fluctuations. For this system,
energy conservation reads as the first law of thermodynamics:
\begin{equation}
W_\tau = Q_\tau + \Delta U_\tau .
\label{4}
\end{equation}
Here $W_\tau$ is the total work done on the system during time $\tau$,
i.e., pulling it with a constant velocity $\mathbf v^*$ through the
fluid over a time $\tau$; $Q_\tau$ is the heat developed in the water
due to the friction of the Brownian particle during time $\tau$ and
$\Delta U_\tau=U_\tau(\mathbf x_\tau)-U_0(\mathbf x_0)$ is the
potential energy difference of the particle in the harmonic potential
in time $\tau$.  Equation \eq{4} clearly shows the difference between
work and heat in this system.  Physically the pulling of the harmonic
potential drags the Brownian particle along, but with a delay, because
of its friction with the water, while at the same time this delay
necessitates a change in its potential energy from its initial
position $\mathbf x_0$ to its final position $\mathbf x_\tau$. A
non-equilibrium stationary state will be reached when the friction
force cancels the attractive force on the particle due to the harmonic
potential (cf. fig. \ref{fig1}) and the fluctuations around this state
will be studied.

In the non-equilibrium stationary state the averages of the
thermodynamic quantities $\overline{W_\tau}$ and $ \overline{Q_\tau}$
are equal, since $\overline{\Delta U_\tau} = \overline{U_{t+\tau}} -
\overline{U_t} = 0$ then. However, unlike in the cases of pure
(deterministic or stochastic) dynamics, $P(Q_\tau) \neq P(W_\tau)$,
which is the main topic of this paper.

In the following two sections we first sketch how the distribution
functions for $W_\tau, Q_\tau$ and $\Delta U_\tau$ for such a system
are obtained, after which the fluctuation theorem, involving the
ratios of the probabilities of $W_\tau$ and $-W_\tau$, and of $Q_\tau$
and $-Q_\tau$ will be stated.

\section{Distribution functions}

\subsection{Work}

The probability distribution function of the work $W_\tau$ done on the
above described system during a time $\tau$ can be derived directly
from the Langevin equation \cite{VanZonCohen02b}.  To do that, note
first that since the Langevin equation \eq{2} is linear in $\mathbf
x_t$, the distribution function of $\mathbf x_t$ is Gaussian. (It is
an Ornstein-Uhlenbeck process.) Secondly, since $ W_\tau = - \mathbf
v^*\cdot \int^\tau_0 [\mathbf x_t - \mathbf x^*_t] \,dt$ depends also
linearly on $\mathbf x_t$, its distribution is Gaussian too.
Therefore it is completely determined by its first and second moments
and can be written in the form:
\begin{eqnarray}
   P(W_\tau) \sim e^{-(W_\tau-\overline{W_\tau})^2/2\sigma^2},
\label{5}
\end{eqnarray}
where the variance $\sigma^2 =
\overline{(W_\tau-\overline{W}_\tau)^2}$. Here the averages are over
all time segments of the trajectory of duration $\tau$. The first and
second moment can be computed from the solution of the Langevin
equation \eq{2} \cite{VanZonCohen02b}.

\subsection{Heat}
\label{heatdist}

\begin{figure}[t]
\centerline{\includegraphics[width=0.62\textwidth]{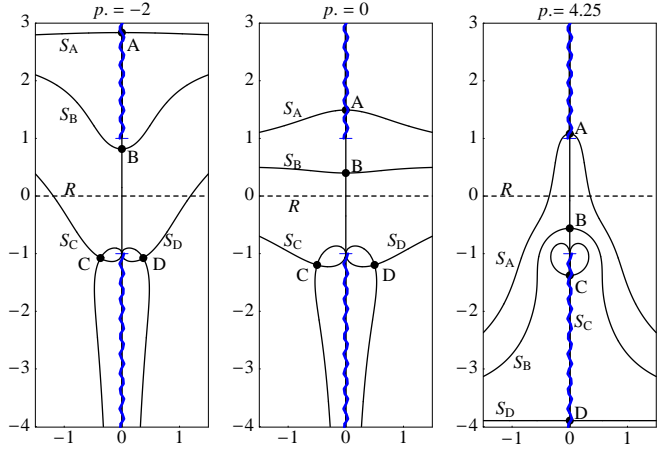}}
\caption{Structure of the complex function $\hat{P}_\tau(q)e^{-iq
  Q_\tau}$ in the complex plane, with $p=Q_\tau/\overline{Q_\tau}=-2$,
  $0$ and $4.25$, respectively.  The saddle points A, B, C and D are
  shown as dots. Wiggly curves along the imaginary axis are branch
  cuts, ending in the branch points $q_1$ (top cut) and $q_2$ (bottom
  cut) indicated with a thin horizontal line.  The dashed line $R$ is
  the real axis and $S_{\mathrm A}$, $S_{\mathrm B}$, $S_{\mathrm C}$
  and $S_{\mathrm D}$ are paths of steepest descent though the saddle
  points, at which the function attains a maximum.}
  \label{fig2}
\end{figure}

A similar simple direct derivation of $P(Q_\tau)$ as was done for
$P(W_\tau)$ cannot be performed, since the $Q_\tau$ given by
(cf. eq.~\eq{4}):
\begin{equation}
  Q_\tau = W_\tau - \Delta U_\tau
\label{4p}
\end{equation}
is quadratic in $\mathbf x_t$ via the $\Delta U_\tau$
(cf. eq.~\eq{3}).  A way to obtain nevertheless $P(Q_\tau)$ is a much
more complicated procedure via its Fourier transform $\hat{P}_\tau
(q)$:
\begin{equation}
 P(Q_\tau) = \frac{1}{2\pi} \int^\infty_{-\infty} \!dq\,
 \hat{P}_\tau(q) e^{-iq Q_\tau}.
\label{6}
\end{equation}
$\hat{P}_\tau (q)$ can be computed exactly for all $\tau$ (see
ref.~\cite{VanZonCohen03a,VanZonCohen03b}):
\begin{eqnarray}
 \hat{P}_\tau (q) & = & \int^\infty_{-\infty}\! dQ_\tau\,  P(Q_\tau)
 e^{iq Q_\tau}
 =\frac{ \exp \left[q(i-q)v^{*^{2}} \{
  \tau-\frac{2q^2(1-e^{-\tau})^2} {1+(1-e^{-2\tau})q^2}\}\right] }
  {[1+(1-e^{-2\tau})q^2]^{3/2} }.
\label{7}
\end{eqnarray}
Note that $\hat{P}_\tau(q)$ is a function in the complex plane with
branch cuts (due to the square root in its denominator) and two
singularities at $q_1=i(1-e^{-2\tau})^{-1/2}$ and $q_2=-q_1$, which
are also the end points of the branch cuts from $q_1$ to $+i \infty$
and $q_2$ to $-i \infty$ (cf.\ fig.\ \ref{fig2}).  All singularities
occur for imaginary $q$-values and introduce exponential (rather than
Gaussian) tails in $P(Q_\tau)$ for large $Q_\tau$.

The approximate evaluation of $P(Q_\tau)$ can be performed for large
$\tau$ using the Saddle Point Method (SPM) \cite{Jeffreys},
capitalizing on the fact that $Q_\tau$ in the exponent on the
right-hand side of eq.~\eq{6} is proportional to $\tau$. As seen in
fig.~\ref{fig2}, the integral $\int^\infty_{-\infty}$ along the real
axis $R$ in eq.~\eq{6} can, for every $Q_\tau$, be deformed to a path
of steepest descent $S_{\mathrm{B}}$ in the complex plane that goes
through the saddle point B, without passing through a singularity ---
which is not possible for the saddle points A, C and D. All that is
needed to evaluate the integral along the real axis are then the
properties of the function at the point B. For details we refer to a
forthcoming paper \cite{VanZonCohen03b}.

To summarize the results of the distribution functions for the
fluctuations of all three quantities occurring in eq.~\eq{4}:

1. $P(W_\tau) \sim e^{-(W_\tau-\overline{W}_\tau)^2/2\sigma^2}$ is
Gaussian, from the Langevin eq.~\eq{5},

2. $P(\Delta U_\tau) \sim e^{-\beta|\Delta U_\tau|}$ is exponential.
This can be physically understood by observing that for large $\tau,
\Delta U_\tau$ is the difference of two independent quantities,
distributed as the potential energy of a (Brownian) particle in a
potential in contact with a heat bath, i.e., with a Boltzmann weight
$e^{-\beta U}$.

\begin{figure}[t]
  \centerline{\includegraphics[width=0.6\textwidth]{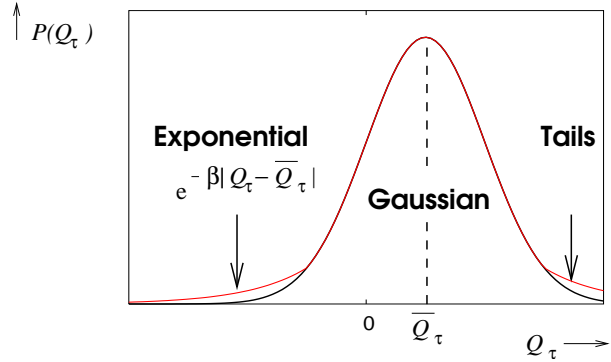}}
  \caption{Mixed Gaussian and exponential behavior of $P(Q_\tau)$ (sketch).}
  \label{fig4}
\end{figure}

3. Since $Q_\tau = W_\tau -\Delta U_\tau$, $P(Q_\tau)$ results from an
interplay of the Gaussian $P(W_\tau)$ and the exponential
$e^{-\beta\Delta U(\tau)}$. Thus the SPM leads to a mixed curve for
$P(Q_\tau)$ which is Gaussian-like in the center, but has exponential
tails.  (fig.~\ref{fig4}).

\section{Fluctuation Theorems}
\subsection{Work}
\label{work}
We start by introducing a new formulation of the FT for $P(W_\tau)$
which is more precise than that given (for $P(Q_\tau)$) in
eq.~\eq{1}. Taking the logarithm of both sides of eq.~\eq{1}, with
$W_\tau$ instead of $Q_\tau$, and dividing both sides then by
$\overline{W}_\tau$, we obtain a reformulation of eq.~\eq{1} of the
form:
\begin{equation}
\lim_{\tau \rightarrow \infty} F_\tau (p_W) = p_W
\label{8}
\end{equation}
where $F_\tau$ is the fluctuation function
\begin{equation}
F_\tau(p_W) = \frac{1}{\overline{W}_\tau} \ln
\frac{P(W_\tau)}{P(-W_\tau)}
\label{9}
\end{equation}
and $p_W = W_\tau /\overline{W}_\tau$ is a scaled work
fluctuation. Equation \eq{8} expresses then the CFT for $W_\tau$,
proven in \cite{VanZonCohen02b}. A new FT for {\it finite} $\tau$ for
the work can be obtained from the Langevin equation \eq{2} and
eq.~\eq{5}, of the form \cite{VanZonCohen02b}:
\begin{equation}
F_\tau(p_W) = \frac{p_W}{1-\varepsilon (\tau)} \approx p_W +
O\left(\frac{1}{\tau}\right)
\label{10}
\end{equation}
where $\varepsilon(\tau)=(1-e^{-\tau})/\tau$.  We note that all CFTs
used so far in the literature restrict themselves to $\tau \rightarrow
\infty$.  In this model one can also discuss the finite $\tau$
behavior and its correction to the infinite time behavior, which shows
that the slope of $F_\tau(p_W)>1$ for all finite $\tau$
(cf. fig.~\ref{fig5}).

\subsection{Heat}
\label{heat}

Proceeding similarly with eq.~\eq{1} for $P(Q_\tau)$ as in \ref{work}
for $P(W_\tau)$ one finds the CFT in a more satisfactory form:
\begin{equation}
\lim_{\tau \rightarrow \infty} F_\tau (p_Q) = p_Q
\label{11}
\end{equation}
where $p_Q = Q_\tau / \overline{Q_\tau}$ is a scaled heat fluctuation,
and 
\begin{equation}
F_\tau(p_Q) = \frac{1}{\overline{Q}_\tau} \ln
\frac{P(Q_\tau)}{P(-Q_\tau)}
\label{9p}
\end{equation}
Contrary to eq.~\eq{8} for $W_\tau$, the relation in eq.~\eq{11} is in
fact incorrect, due to the interaction of $P(Q_\tau)$ with the
(exponential) $P(\Delta U_\tau)$.

\begin{figure}[t]
  \centerline{\includegraphics[angle=-90,width=0.55\textwidth]{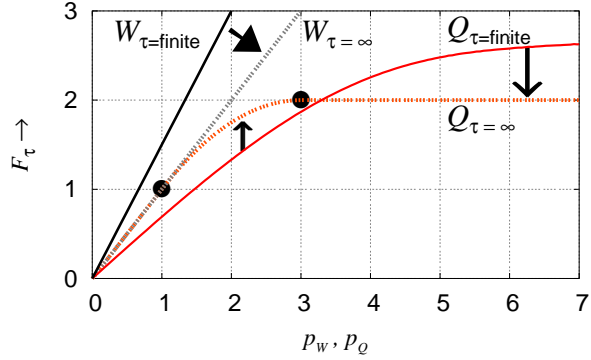}}
  \caption{CFT and NFTs for work and heat fluctuations.}
  \label{fig5}
\end{figure}

A NFT can be derived using the SPM \cite{Jeffreys}, both for infinite
and for finite~$\tau$.  The behavior is determined by the above
mentioned singularities in the complex plane in carrying out the SPM.
The result for $F_\tau(p_Q)$ versus $p_Q$ is given in fig.~\ref{fig5},
together with that of $F_\tau(p_W)$ versus $p_W$ for comparison.

While $F_\tau(p_W)$ versus $p_W$ is linear for all $\tau$, the
behavior of $F_\tau(p_Q)$ versus $p_Q$ is much more complicated. In
fact, for $\tau \rightarrow \infty$ there are three
regimes\cite{VanZonCohen03a}:
\begin{eqnarray}
 F_\tau (p_Q) & = & 
 \left\{\begin{array}{ll}
 p_Q  & {\rm{for}} \;   0 < p_Q < 1 \\
 p_Q - \frac{(1-p_Q)^2}{4} &  {\rm{for}} \;  1 < p_Q < 3\\
 2 & {\rm{for}} \; p_Q > 3
\end{array}\right.
\label{12}
\end{eqnarray}
Thus, for infinite $\tau$, the NFT coincides with the CFT for small
fluctuations $0 < p_Q < 1$, then exhibits a parabolic behavior between
$1 < p_Q <3$ and finally reaches a plateau where $F_\tau(p_Q) = 2$ for
all $p_Q > 3$. The behavior for $p_Q < 0$ follows from the asymmetry
of $F_\tau(p_Q)$ (cf. eq.~\eq{9p}).

The SPM also allows to study analytically the approach of the finite
$\tau$ behavior to the infinite $\tau$ behavior, giving for large but
finite $\tau$:
\begin{eqnarray}
F_\tau(p_Q) & = & \left\{\begin{array}{ll}
p_Q+\frac{h(p_Q)}{\tau} + O\left(\frac{1}{\tau^2}\right) &
{\rm{for}} \; |p_Q| < 1 \\
2 + \frac{g(p_Q)}{\sqrt{\tau}} + O\left(\frac{1}{\tau}\right) &
{\rm{for}} \; p_Q > 3
\end{array}\right.
\label{13}
\end{eqnarray}
where
\begin{eqnarray}
h(p) &=& \frac{8p}{9-p^2} - \frac{3}{2v^{*^{2}}} \ln
\left[\frac{(3-p)(1+p)}{(3+p)(1-p)}\right]
\label{14}
\\
g(p) &=& \sqrt{8(p-3)}.
\label{15}
\end{eqnarray}
Equations \eq{13}--\eq{15} show that the asymptotic behavior of
$F_\tau (p_Q)$ for $p_Q > 3$, is a slowly increasing function $\sim
\sqrt{p_Q-3}$, while the asymptotic $\tau \rightarrow \infty$ curve is
approached as $\tau^{-1/2}$ for $p_Q > 3$ and as $\tau^{-1}$ for $p_Q
< 1$.

\renewcommand\arraystretch{1}
\begin{table}[t]
\begin{center}
\begin{tabular}{|c||c|c|}
\hline
&   Work   &   Heat
\\
\hline\hline
symbol $X_\tau$ 
& 
$W_\tau$ 
& 
$Q_\tau$
\\\hline
distribution 
&   
Gaussian 
& 
Gauss.+Exp. tails
\\\hline
\begin{minipage}{0.6in}
fluctuation \\
function
\end{minipage}
& 
\multicolumn{2}{c|}
{
\raisebox{-3ex}{\rule{0ex}{8ex}}
$\displaystyle F_\tau=\frac{1}{\overline X_\tau}\ln\frac{P(X_\tau)}{P(-X_\tau)}$}
\\\hdashline
$\tau\to\infty$ 
&
\emph{Conventional}
& 
\emph{New}
\\
&
$F_\tau$ straight line with slope $1$
&
$F_\tau$ has slope $1$ for small $Q_\tau$
\\
&
for all $W_\tau$  
&  
$F_\tau=2$ for large $Q_\tau$
\\\hdashline
$\tau$ finite 
&
\emph{New}
& 
\emph{New}
\\
& 
$F_\tau$ straight line with slope $>1$
& 
$F_\tau$ no slope $1$ for small $Q_\tau$
\\
       & for all $W_\tau$          &  
$F_\tau$ increasing for large $Q_\tau$
\\\hdashline
\begin{minipage}{0.6in}
\ \\
plots\\
$F_\tau(p)$\\
versus\\
$p$
\end{minipage}
&
\raisebox{-11ex}{\rule{0ex}{17.5ex}}
\raisebox{25pt}{\includegraphics[angle=-90,width=0.31\textwidth]{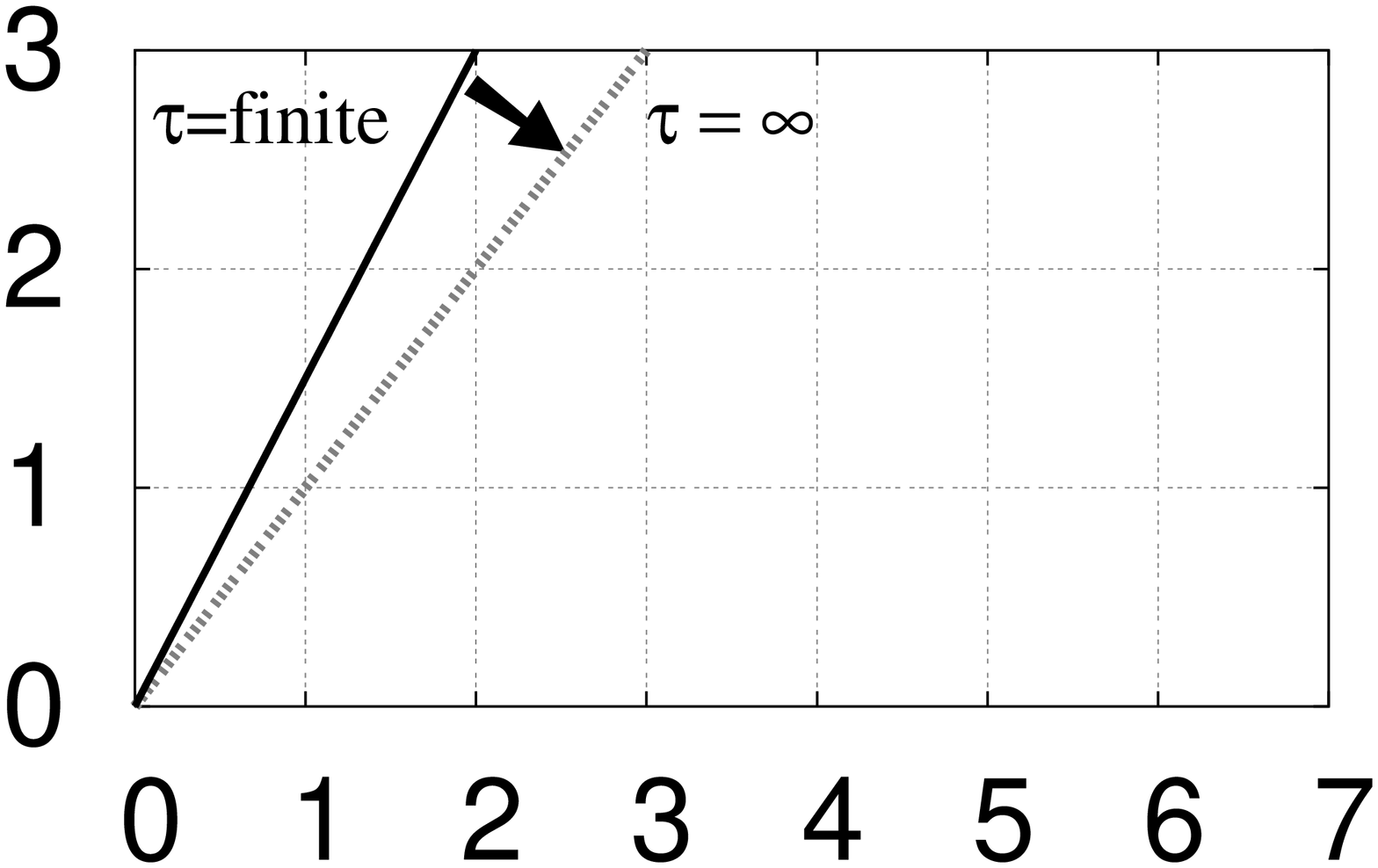}}
&
\raisebox{25pt}{\includegraphics[angle=-90,width=0.31\textwidth]{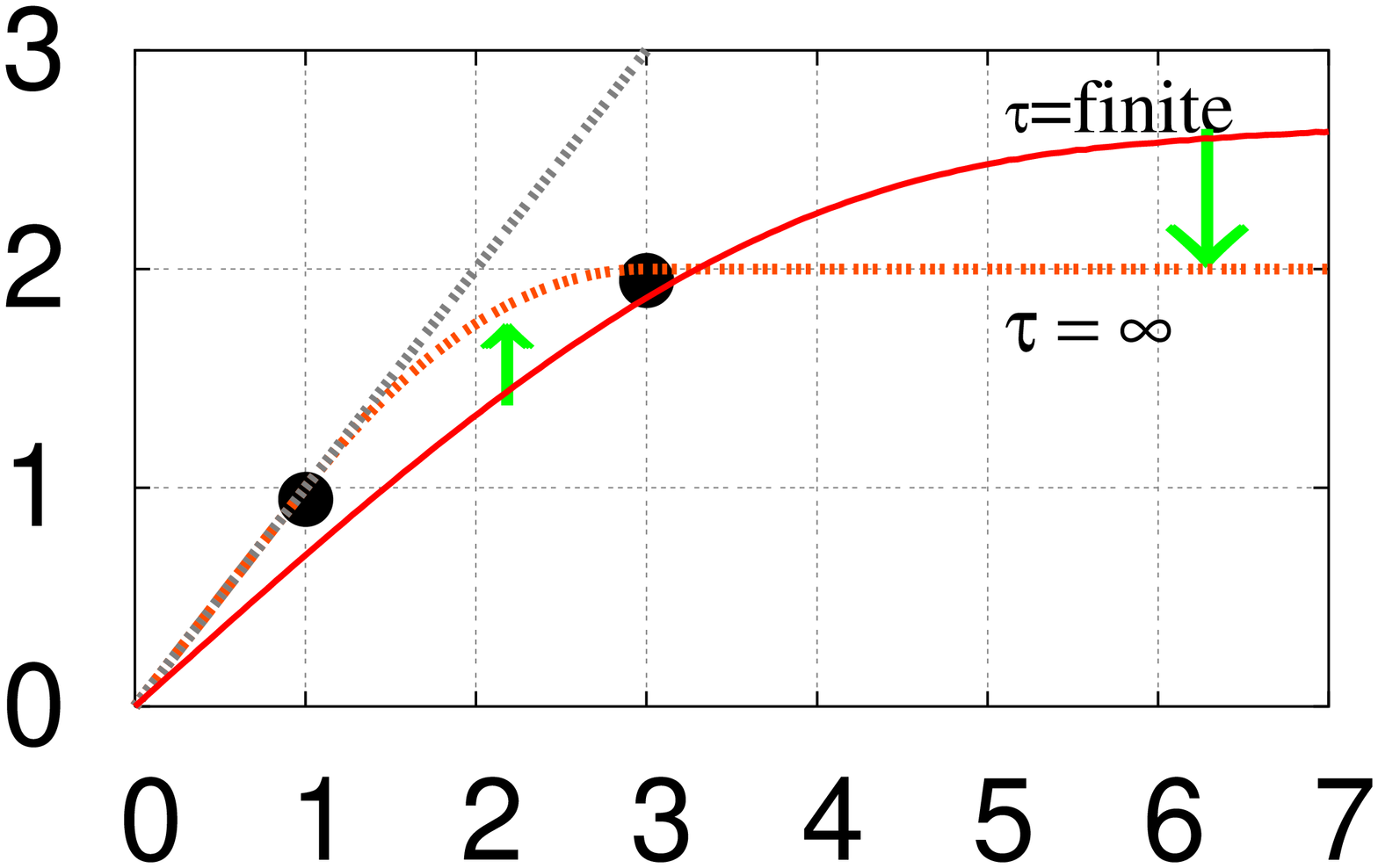}}
\\\hline
\end{tabular}
\end{center}
\ \\
\noindent
Table I. Results for the work 
(section \ref{work}) and heat (section \ref{heat}) fluctuations.
\end{table}

\section{Discussion}

In table I, the results for the CFT and the NFT for work and heat are
summarized.  A number of questions and remarks present themselves.

1. How general is the NFT for systems with mixed deterministic and
stochastic dynamics?  Is the Boltzmann factor and the ensuing
exponential decay for large fluctuations more general than in this
model? One would be inclined to think so, in view of the physical
argument of point 2 in \ref{heatdist}.

2. The relative probability for the Brownian particle to gain rather
than supply heat to the water is much larger in the NFT than in the
CFT (cf. fig.~\ref{fig5}). This might be of relevance in designing
micro or nano machines sensitive in their functioning to the heat
absorbed during large fluctuations.  The CFT would not be a good basis
to judge this effect.  We remark that since $F_\tau(p_Q)>0$ for
$p_Q>0$ in the NFT, $\overline{Q}_{\tau}>0$, in accordance with the
Second Law.

3. The plateau value of 2 for $F(p_Q)$ for large $p_Q$ $(> 3)$ can be
understood physically.  For the probability for large $Q_\tau$
(i.e. $p_Q$), the exponential distribution $P(\Delta U_\tau)\sim
e^{-\beta |\Delta U_\tau|}$ dominates over the Gaussian distribution
$P(W_\tau)$ (cf. fig.~\ref{fig4}). This implies, with
$\overline{W}_\tau=\overline{Q}_\tau$ (see the end of section
\ref{secsystem}), that $P(Q_\tau)\sim e^{-\beta|Q_\tau -
  \overline{Q}_\tau|}$ so that $F_\tau(p_Q)=2$, with $\beta = 1$.

4. Similar results as discussed in this paper are obtained for the
Transient Fluctuation Theorem \cite{Evansetal94,CohenGallavotti99}.
It is certainly, at least in this model, not the identity for all
$\tau$ which obtains in the CFT\cite{VanZonCohen02b,VanZonCohen03b}.

5. All our analytic results have been verified by comparison with the
results of two numerical methods: a sampling method and a fast inverse
Fourier transform of eq.~\eq{6}\cite{VanZonCohen03b}. In particular, the
SPM turns out to give good results already for $\tau > 3$, whereas
curves for $\tau \leq 3$ need to be obtained numerically.

6. The connection between the NFT and theorems of
Irreversible Thermodynamics is unclear, although the linear (CFT-like)
behavior of the NFT for small fluctuations with $0 < p_Q < 1$
(cf. fig.~\ref{fig5}) suggests that the same relations hold as for the CFT
for small deviations near equilibrium.

7. So far, in all cases dealt with here, only one property of the
fluctuations of the thermodynamic quantities work and heat in a
non-equilibrium stationary state --- whose averages are described by
the First and Second Law of Thermodynamics --- has been discussed,
viz. the fluctuation function $F_\tau$. Whether something can be said
about other properties of fluctuations of thermodynamic quantities,
also beyond the linear regime, remains an interesting open question.

\section*{\bf Acknowledgment}

This research was supported by the Office of Basic Engineering of the
US Department of Energy, under grant No.\ DE-FG-02-88-ER13847.


\begin{thebibliography}{10}
\setlength{\parskip}{0ex}
\setlength{\itemsep}{0ex}

\bibitem{Evansetal93}
D.\ J.\ Evans, E.\ G.\ D.\ Cohen and G.\ P.\ Morriss, Phys.\ Rev.\
Lett.\ 71 (1993) 2401.

\bibitem{GallavottiCohen95a} 
G.\ Gallavotti and E.\ G.\ D.\ Cohen, Phys.\ Rev.\ Lett.\ 74 (1995)
2694; J.\ Stat.\ Phys.\ 80 (1995) 931; E.\ G.\ D.\ Cohen, Physica 240
(1997) 43.

\bibitem{Evansetal94}
D.\ J.\  Evans and D.\ J.\  Searles, Phys.\  Rev.\  {E} 50 (1994) 1645.

\bibitem{CohenGallavotti99}
E.\ G.\ D.\  Cohen and G.\  Gallavotti, J.\  Stat.\  Phys.\  96 (1999) 1343.

\bibitem{DeGrootMazur}
S.\ R.\ de Groot and P.\ Mazur, Non-Equilibrium Thermodynamics, Dover
Publications, Inc.\ , New York, 1984.

\bibitem{Kurchan98}
J.\  Kurchan, J.\  Phys.\  {A}, Math.\  Gen.\  31 (1998) 3719.

\bibitem{LebowitzSpohn99}
J.\ L.\  Lebowitz and H.\  Spohn, J.\  Stat.\  Phys.\  95 (1999) 333.

\bibitem{Gallavotti96}
G.\  Gallavotti, Phys.\  Rev.\  Lett.\  77 (1996) 4334.

\bibitem{Wangetal02} 
G.\ M.\ Wang {\it et al.}, Phys.\ Rev.\ Lett.\ 89 (2002) 050601.

\bibitem{VanZonCohen02b}
R.\  van Zon and E.\ G.\ D.\  Cohen, Phys.\  Rev.\  {E} 67 (2003) 046102.

\bibitem{VanZonCohen03a}
R.\  van Zon and E.\ G.\ D.\  Cohen, Phys.\  Rev.\  Lett.\  91 (2003) 110601.

\bibitem{Jeffreys}
H.\ Jeffreys and B.\ S.\ Jeffreys, Methods of Mathematical Physics,
3rd Edition, Cambrige University Press, Cambridge, 1956.

\bibitem{VanZonCohen03b}
R.\  van Zon and E.\ G.\ D.\  Cohen (in preparation).

\end{thebibliography}
\end{document}